\documentclass[prb,aps,twocolumn,amsmath,amssymb,floatfix,showpacs,
superscriptaddress]{revtex4}

\usepackage[dvips]{graphics}
\usepackage{color}
\definecolor{dred}{rgb}{0,0,0.6}

\begin{document}

\title{\textcolor{dred}{Characteristics of persistent spin current components 
in a quasi-periodic Fibonacci ring with spin-orbit interactions: Prediction
of spin-orbit coupling and on-site energy}}

\author{Moumita Patra}

\affiliation{Physics and Applied Mathematics Unit, Indian Statistical
Institute, 203 Barrackpore Trunk Road, Kolkata-700 108, India}

\author{Santanu K. Maiti}

\email{santanu.maiti@isical.ac.in}

\affiliation{Physics and Applied Mathematics Unit, Indian Statistical
Institute, 203 Barrackpore Trunk Road, Kolkata-700 108, India}

\begin{abstract}

In the present work we investigate the behavior of all three components 
of persistent spin current in a quasi-periodic Fibonacci ring subjected
to Rashba and Dresselhaus spin-orbit interactions. Analogous to persistent
charge current in a conducting ring where electrons gain a Berry phase 
in presence of magnetic flux, spin Berry phase is associated during 
the motion of electrons in presence of a spin-orbit field which is 
responsible for the generation of spin current. The interplay between 
two spin-orbit fields along with quasi-periodic Fibonacci sequence on 
persistent spin current is described elaborately, and from our analysis,
we can estimate the strength of any one of two spin-orbit couplings together
with on-site energy, provided the other is known. 

\end{abstract}


\maketitle

\section{Introduction}

The study of spin dependent transport in low-dimensional quantum systems,
particularly ring-like geometries, has always been intriguing due to their 
strange behavior and possible potential applications in designing spintronic
devices. To do this proper spin regulation is highly important. Several 
attempts~\cite{dev1,th1,th2,th3,th4,th5,th6,th7,th8,th9,th10} have been 
made in the last couple of decades, and undoubtedly, a wealth of literature 
knowledge has been established towards this direction. Mostly external 
magnetic fields or ferromagnetic leads were used~\cite{conv1,conv2} to 
control electron spin but none of these are quite suitable from experimental 
perspective. This is because confining a large magnetic field in a small
\begin{figure}[ht]
{\centering \resizebox*{4.5cm}{3.5cm}{\includegraphics{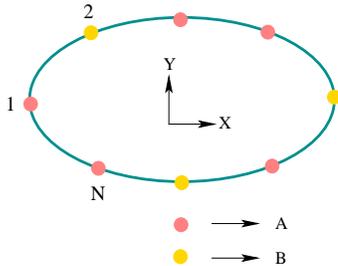}}\par}
\caption{(Color online). Schematic view of a $8$-site mesoscopic Fibonacci 
ring ($5$th generation) subjected to Rashba and Dresselhaus spin-orbit 
interactions. The ring is constructed with two basic atomic units 
$A$ and $B$, and they are described by two colored filled circles.}
\label{Model}
\end{figure}
region e.g., ring-like geometry is always a difficult task, and at the same
footing major problem arises during spin injection from ferromagnetic 
leads through a conducting junction due to large inconsistency in resistivity.
Certainly it demands new methodologies and attention has been paid towards 
the intrinsic properties~\cite{so1,so2,so3,so4,so5,so6,so7,so8,so9,so10,n1,n2} 
like spin-orbit (SO) coupling of the materials. Usually two types of SO fields 
are encountered in studying spin dependent transport phenomena. One is known 
as Rashba SO coupling~\cite{rashba} which is generated due to 
breaking of symmetry in confining potential, and thus, can be regulated 
externally~\cite{guo} by gate electrodes. While the other, defined as 
Dresselhaus SO coupling~\cite{dressel}, is observed due to the breaking 
of structural symmetry. The SO field plays an essential role in spintronic 
applications as it directly couples to the electron's spin degree of freedom.

In presence of such SO field a net circulating {\em spin current} is 
established~\cite{qf1} in a conducting ring, analogous to magnetic flux driven
persistent {\em charge current}~\cite{ab1,ab2,ab3}. The magnetic flux 
introduces a phase, called Berry phase, to moving electrons which produces 
net charge current by breaking time reversal symmetry between clockwise and 
anti-clockwise moving electrons, while a spin Berry phase is associated in 
presence of SO coupling which generates spin current. 

The works involving persistent spin current in ring-like geometries studied 
so far are mostly confined to the perfect periodic lattices or 
completely random ones~\cite{ding,AC,san1,san2}. But, to the best of our 
knowledge, no one has addressed the behavior of SO-interaction induced spin 
current in quasi-periodic lattices which can bridge the gap between an 
ordered lattice and a fully random one. In addition, the earlier studies 
essentially focused on only one component (viz, $Z$-component), though it 
is extremely 
interesting and important too to study all three components of spin 
current to analyze spin dynamics of moving electrons in presence of SO
fields. Motivated with this, in the present work we explore the behavior 
of persistent spin current in a one-dimensional (1D) quasi-periodic ring 
geometry where lattice sites are arranged in a Fibonacci 
sequence~\cite{site1,site2,site3}, the simplest example of a quasi-periodic 
system. A Fibonacci chain is 
constructed by two basic units $A$ and $B$ following the specific rule
$A \rightarrow AB$ and $B \rightarrow A$. Thus, applying successively this
substitutional rule, staring from $A$ lattice or $B$ lattice we can 
construct the full lattice for any particular generation, say $p$-th 
generation, obeying the prescription $F_p=F_{p-1} \otimes F_{p-2}$, and
connecting its two ends we get the required ring model. Thus, if we start
with $A$ lattice then $A$, $AB$, $ABA$, $ABAAB$, $ABAABABA$, $\dots$, etc.,
are the first few generations, and the series is characterized by the ratio
of total number of $A$ and $B$ atoms which is called as Golden mean $\tau$
($=1+\sqrt{5}/2$). Now, instead of considering $A$ and $B$ as lattice points 
if we assign them as bond variables then {\em bond model} of Fibonacci 
generation is established~\cite{bond1,bond2}, and when both these lattice 
and bond models are taken into account it becomes a 
{\em mixed model}~\cite{site2}. In our model we consider only site 
representation of Fibonacci sequence starting with lattice $A$, for the 
sake of simplification. The response of other will be discussed elsewhere 
in our future work.

In this work, we address the behavior of all three components of persistent
spin current in a Fibonacci ring subjected to Rashba and Dresselhaus SO
couplings. Within a tight-binding framework we calculate the current 
components using second-quantized approach which is the most convenient 
tool for such calculations. The interplay between Rashba and Dresselhaus 
SO fields on the current components exhibits several interesting patterns
that can be utilized to estimate any one of the SO fields if we know the 
other one, and also we can estimate the site energy of $A$- or $B$-type 
site provided any one of these two is given. Our analysis can be utilized 
to explore spin dependent phenomena in any correlated lattices subjected 
to such kind of SO fields. This is the first step towards this direction.

The work is arranged as follows. In Sec. II we present the model and 
theoretical framework for calculations. The results are discussed in 
Sec. III, and finally we conclude in Sec. IV.

\section{Model and Theoretical Formulation}

The mesoscopic Fibonacci ring subjected to Rashba and Dresselhaus SO
couplings is schematically depicted in Fig.~\ref{Model}, and the
Hamiltonian of such a ring reads as,
\begin{equation}
\mbox{\boldmath $H$}=\mbox{\boldmath $H$}_{\mbox{\tiny 0}} +
\mbox{\boldmath $H$}_{\mbox{\tiny R}} + \mbox{\boldmath $H$}_{\mbox{\tiny D}}
\label{eq1}
\end{equation}
which includes the SO-coupled free term 
($\mbox{\boldmath $H$}_{\mbox{\tiny 0}}$), Rashba Hamiltonian 
($\mbox{\boldmath $H$}_{\mbox{\tiny R}}$) and the Dresselhaus Hamiltonian
($\mbox{\boldmath $H$}_{\mbox{\tiny D}}$). Using tight-binding (TB) 
framework we can express these Hamiltonians~\cite{site3} for a $N$-site 
ring as follows:
\begin{subequations}
\begin{align}
\mbox{\boldmath $H$}_{\mbox{\tiny 0}} &= &\sum\limits_{n} \mbox{\boldmath $c$}
_n^\dagger \mbox{\boldmath $\epsilon$}_n \mbox{\boldmath $c$}_n +
\sum\limits_{n}\left( \mbox{\boldmath $c$}_{n+1}^\dagger\mbox{\boldmath $t$}
\mbox{\boldmath $c$}_n + \mbox{\boldmath $c$}_{n}^\dagger\mbox{\boldmath $t$}
\mbox{\boldmath $c$}_{n+1} \right) \\
\mbox{\boldmath $H$}_{\mbox{\tiny R}} & = &-\sum\limits_{n}\left(
\mbox{\boldmath $c$}_{n+1}^\dagger\left(i \mbox{\boldmath $\sigma$}_x\right)
\mbox{\boldmath $\alpha$} \cos \phi_{n,n+1} \mbox{\boldmath $c$}_n +
h.c.\right)\nonumber \\
 & & - \sum\limits_{n}\left(\mbox{\boldmath $c$}_{n+1}^\dagger\left(i
\mbox{\boldmath $\sigma$}_y\right)\mbox{\boldmath $\alpha$} \sin\phi_{n,n+1}
\mbox{\boldmath $c$}_n + h.c.\right) \\
\mbox{\boldmath $H$}_{\mbox{\tiny D}} &  = & \sum\limits_{n}\left(
\mbox{\boldmath $c$}_{n+1}^\dagger\left(i\mbox{\boldmath $\sigma$}_y\right)
\mbox{\boldmath $\beta$}\cos\phi_{n,n+1} \mbox{\boldmath $c$}_n + h.c.\right)
\nonumber \\
 & & + \sum\limits_{n}\left(\mbox{\boldmath $c$}_{n+1}^\dagger
\left(i\mbox{\boldmath $\sigma$}_x\right) \mbox{\boldmath $\beta$}
\sin\phi_{n,n+1}\textbf{c}_n + h.c.\right)
\end{align}
\label{hamil}
\end{subequations}
\vskip 0.01cm
\noindent
where the site index $n$ runs from $1$ to $N$, and we use the condition
$N+1=1$. The other factors are: 
\vskip 0.2cm
\noindent
$\mbox{\boldmath $\epsilon$}_n=\left(\begin{array}{cc}
    \epsilon_{n\uparrow} & 0 \\ 
    0 & \epsilon_{n\downarrow}
\end{array}\right)$, $\mbox{\boldmath $t$}=\left(\begin{array}{cc}
    t & 0 \\ 
    0 & t
\end{array}\right)$, 
$\mbox{\boldmath $c$}_n=\left(\begin{array}{cc}
    c_{n\uparrow} \\ c_{n\downarrow}
\end{array}\right)$, 
\vskip 0.3cm
\noindent
$\mbox{\boldmath $\alpha$}=\left(\begin{array}{cc}
    \alpha & 0 \\ 
    0 & \alpha
\end{array}\right)$, $\mbox{\boldmath $\beta$}=\left(\begin{array}{cc}
    \beta & 0 \\ 
    0 & \beta
\end{array}\right)$ 
\vskip 0.3cm
\noindent
where $t$ is the nearest-neighbor hopping integral and $\epsilon_{n\uparrow}$ 
($\epsilon_{n\downarrow}$) represents the on-site energy of an up (down) spin 
electron sitting at the $n$th site. As we are not considering any magnetic 
type interaction the site energies for up and down spin electrons become
equal i.e., $\epsilon_{n\uparrow}=\epsilon_{n\downarrow}=\epsilon_n$ (say). 
For the $A$-type sites we refer $\epsilon_n=\epsilon_A$, and similarly,
$\epsilon_n=\epsilon_B$ for $B$-type atomic sites. When these two site
energies are identical (viz, $\epsilon_A=\epsilon_B$) the Fibonacci ring
becomes a perfect one, and in that case we set them to zero without loss of
any generality. $\alpha$ and $\beta$
are the Rashba and Dresselhaus SO coupling strengths, respectively, and
$\phi_{n,n+1}=\left(\phi_n+\phi_{n+1}\right)/2,$ where $\phi_n=2\pi(n-1)/N$.
$\mbox{\boldmath $\sigma$}_i$'s ($i=x$, $y$, $z$) are the Pauli spin 
matrices in $\mbox{\boldmath $\sigma$}_z$ diagonal representation.

Looking carefully the Rashba and Dresselhaus Hamiltonians (Eqs.~\ref{hamil}(b)
and \ref{hamil}(c)) one can find that these two Hamiltonians are connected
by a unitary transformation 
$\mbox{\boldmath $H$}_{\mbox{\tiny D}}=\mbox{\boldmath $U$}
\mbox{\boldmath $H$}_{\mbox{\tiny R}} \mbox{\boldmath $U$}^{\dagger}$, 
where $\mbox{\boldmath $U$}=\mbox{\boldmath $\sigma$}_z 
(\mbox{\boldmath $\sigma$}_x + \mbox{\boldmath $\sigma$}_y)/\sqrt{2}$.
This hidden transformation relation leads to several interesting results,
in particular when the strengths of these two SO couplings are equal, which
will be available in our next section (Sec. III).

To calculate spin current components we define the operator~\cite{ding} 
$\mbox{\boldmath $I$}_k=\frac{1}{2N}\left(\mbox{\boldmath $\sigma$}_k
\dot{\mbox{\boldmath $x$}} + \dot{\mbox{\boldmath $x$}}
\mbox{\boldmath $\sigma$}_k\right)$, where $k=x$, $y$, $z$ depending 
on the specific component. In this expression $\dot{\mbox{\boldmath $x$}}$
is obtained by taking the commutation of position operator
$\mbox{\boldmath $x$}$ ($=\sum\limits_{n}\mbox{\boldmath $C$}_n^\dagger
n \mbox{\boldmath $C$}_n$) with the Hamiltonian $\mbox{\boldmath $H$}$.
After simplification the current operator gets the form:
\begin{eqnarray}
\mbox{\boldmath $I$}_k & = & \frac{i\pi}{N}\sum\limits_{n} \left(
\mbox{\boldmath $c$}_n^\dagger \mbox{\boldmath $\sigma$}_k\mbox{\boldmath $t$}
_{\phi}^{\dagger n,n+1} \mbox{\boldmath $c$}_{n+1}- h.c. \right) \nonumber \\
 & & + \frac{i\pi}{N}\sum\limits_{n} \left( \mbox{\boldmath
$c$}_n^\dagger \mbox{\boldmath $t$}_{\phi}^{\dagger n,n+1} 
\mbox{\boldmath $\sigma$}_k \mbox{\boldmath $c$}_{n+1}- h.c. \right) 
\label{eq4}
\end{eqnarray}
where, $\mbox{\boldmath $t$}_{\phi}^{n,n+1}$ is a ($2\times2$) matrix 
whose elements are 
\vskip 0.3cm
\noindent
${\mbox{\boldmath $t$}_{\phi}^{n,n+1}}_{1,1}=t$ 
\vskip 0.2cm
\noindent
${\mbox{\boldmath $t$}_{\phi}^{n,n+1}}_{2,2}=t$
\vskip 0.2cm
\noindent
${\mbox{\boldmath $t$}_{\phi}^{n,n+1}}_{1,2}=-i\alpha e^{-i\phi_{n,n+1}}
+ \beta e^{i \phi_{n,n+1}}$
\vskip 0.2cm
\noindent
${\mbox{\boldmath $t$}_{\phi}^{n,n+1}}_{2,1}=
-i\alpha e^{i \phi_{n,n+1}}-\beta e^{i\phi_{n,n+1}}$.
\vskip 0.3cm
\noindent
Once the current operator is established (Eq.~\ref{eq4}), the individual 
current components carried by each eigenstate, say $|\psi_m\rangle$, can 
be easily found from the operation 
$I_{k,m}=\langle\psi_m|\mbox{\boldmath $I$}_k|\psi_m\rangle$, where
$|\psi_m\rangle=\sum\limits_{n}\left(a_{n\uparrow}^m
|n\uparrow\rangle + a_{n\downarrow}^m|n \downarrow \rangle\right)$.
$a_{n\sigma}^m$'s are the coefficients. Doing a quite long and 
straightforward calculation we eventually get the following current 
expressions for three different directions ($X$, $Y$ and $Z$) as follows:
\begin{subequations}
\begin{align}
I_{z,m} & = & \frac{2\pi it}{N}\sum\limits_{n}\left[
\left\{a_{n,\uparrow}^{m\ast} a_{n+1,\uparrow}^m - h.c. \right\} 
- \left\{a_{n,\downarrow}^{m\ast} a_{n+1,\downarrow}^m - \right.\right. 
\nonumber \\
 & & \left. \left. h.c. \right\} \right] \\
I_{x,m} & = & \frac{2\pi it}{N}\sum\limits_{n}\left\{ \left( a_{n,\uparrow}^
{m\ast}a_{n+1,\downarrow}^m + a_{n,\downarrow}^{m\ast}a_{n+1,\uparrow}^m
\right) - h.c. \right\}\nonumber \\
 & & + \frac{2\pi}{N}\sum\limits_{n}\left\{ \left( a_{n+1,\uparrow}^{m\ast}
a_{n,\uparrow}^m + a_{n+1,\downarrow}^{m\ast}a_{n,\downarrow}^m \right) + h.c.
\right\} \times \nonumber \\
 & & \left(\beta\sin\phi_{n,n+1}-\alpha\cos\phi_{n,n+1}\right) \\
I_{y,m} & = & \frac{2\pi t}{N}\sum\limits_{n}\left\{ \left( a_{n+1,\downarrow}
^{m\ast}a_{n,\uparrow}^m - a_{n+1,\uparrow}^{m\ast}a_{n,\downarrow}^m \right)
+ h.c. \right\}\nonumber \\
 & & + \frac{2\pi}{N}\sum\limits_{n}\left\{ \left( a_{n+1,\uparrow}^{m\ast}
a_{n,\uparrow}^m + a_{n+1,\downarrow}^{m\ast}a_{n,\downarrow}^m \right) + h.c.
\right\} \times \nonumber \\
 & & \left(\beta\cos\phi_{n,n+1}-\alpha\sin\phi_{n,n+1}\right)
\end{align}
\label{compcurr}
\end{subequations}
Thus, at absolute zero temperature, the net current for a $N_e$ electron
system becomes
\begin{equation}
I_k=\sum\limits_{m=1}^{N_e}I_{k,m}.
\label{eq6a}
\end{equation}

\section{Results}

Based on the above theoretical framework now we present our numerical 
results which include three different current components carried by 
individual energy eigenstates, net currents of all three components for
a particular electron filling $N_e$, and the possibilities of determining
any one of the SO fields as well as on-site energies provided the other is
known. As we are not focusing on quantitative analysis considering a 
particular material, we choose $c=e=h=1$ for the sake of simplification 
and fix the nearest-neighbor hopping integral $t=1.5$ throughout the 
analysis. The values of other parameters are given in subsequent figures.

\subsection{Current components carried by distinct energy levels}

Before analyzing net current components for a specific filling factor,
let us first focus on the behavior of individual state currents as they
give more clear conducting signature of separate energy levels which 
essentially govern the net response of any system.

\begin{figure}[ht]
{\centering \resizebox*{8cm}{8cm}{\includegraphics{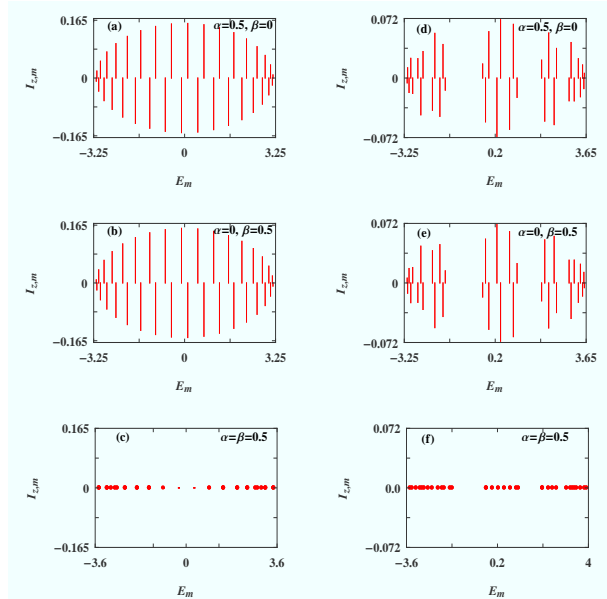}}\par}
\caption{(Color online). $Z$-component of spin current ($I_{z,m}$) carried
by individual energy levels (levels are indexed by the parameter $m$ and 
$E_m$ being the energy eigenvalue of $m$th level) for $34$-site ring ($8$th
generation) considering different values of $\alpha$ and $\beta$, where the
first and second columns correspond to $\epsilon_A=\epsilon_B=0$ and
$\epsilon_A=-\epsilon_B=1$, respectively.}
\label{fig2a}
\end{figure}

In Fig.~\ref{fig2a} we present the variation of $Z$-component of spin 
current carried by individual energy levels for a completely perfect 
(left column) and a Fibonacci ring (right column) in presence of different
SO couplings. Several interesting features are observed. (i) In presence 
of any one of the two SO fields individual state currents exhibit a nice 
pattern for the perfect case where the current starts 
increasing (each vertical line represents the current amplitude for each
state) from the energy band edge and reaches to a maximum at the band
centre. While, for the Fibonacci ring sub-spectra with finite gaps, 
associated with energy sub-bands, are obtained which is the generic feature
of any Fibonacci lattice, like other quasi-crystals~\cite{g1,g2}. In each 
sub-band higher 
currents are obtained from central energy levels while edge states provide
lesser currents, similar to a perfect ring. Thus, setting the Fermi energy, 
associated with electron filling $N_e$, at the centre or towards the edge 
of each sub-band one can regulate current amplitude and this phenomenon
can be visualized at multiple energies due to the appearance of multiple
energy sub-bands. Here it is also crucial to note that no such 
phenomenon will be observed in a completely random disordered ring as 
it exhibits only localized states for the entire energy band region. 
(ii) Successive energy levels carry currents in opposite directions which 
gives an important conclusion that the net $Z$-component
of current is controlled basically by the top most filled energy level,
similar to that what we get in the case of conventional magnetic flux
induced persistent charge current in a conducting mesoscopic ring. 
(iii) Finally, it is important to see that the direction of individual 
state currents in a ring subjected to only Rashba SO coupling gets exactly
reversed when the ring is described with only Dresselhaus SO interaction,
without changing any magnitude. Thus, if both these two SO fields are
present in a particular sample and if they are equal in magnitude then
current carried by different eigenstates drops exactly to zero due to
mutual cancellation of current caused by these fields. This vanishing
nature of spin current can be proved as follows. It is 
already pointed out that $\mbox{\boldmath $H$}_{\mbox{\tiny R}}$ and 
$\mbox{\boldmath $H$}_{\mbox{\tiny D}}$ are connected by a unitary 
transformation $\mbox{\boldmath $H$}_{\mbox{\tiny D}} =
\mbox{\boldmath $U$} \mbox{\boldmath $H$}_{\mbox{\tiny R}} 
\mbox{\boldmath $U$}^{\dagger}$. Therefore, if $|\psi_m\rangle$ be an
eigenstate of $\mbox{\boldmath $H$}_{\mbox{\tiny R}}$ then
$\mbox{\boldmath $U$} |\psi_m\rangle$ ($=|\psi_m^{\prime}\rangle$) will
be the eigenstate of the Hamiltonian $\mbox{\boldmath $H$}_{\mbox{\tiny D}}$.
This immediately gives us the following relation:
\begin{eqnarray}
I_{z,m}|_{\mbox{\tiny D}} & = &
\langle\psi_m^{\prime}|\mbox{\boldmath $I$}_{z,m}|\psi_m^{\prime}\rangle
\nonumber \\
 &=& \langle\psi_m|\mbox{\boldmath $U$}^\dagger\mbox{\boldmath$I$}_{z,m}
\mbox{\boldmath $U$}|\psi_m\rangle\nonumber \\
 &=& \langle\psi_m|\mbox{\boldmath $U$}^\dagger\frac{1}{2N}
\left(\mbox{\boldmath$\sigma$}_z\dot{\mbox{\boldmath $x$}} + 
\dot{\mbox{\boldmath $x$}} \mbox{\boldmath $\sigma$}_z\right)
\mbox{\boldmath $U$}|\psi_m\rangle\nonumber \\
 &=& \langle\psi_m|\frac{1}{2N}\left(-\mbox{\boldmath$\sigma$}_z
\dot{\mbox{\boldmath $x$}} - \dot{\mbox{\boldmath $x$}}
\mbox{\boldmath $\sigma$}_z\right)|\psi_m\rangle\nonumber \\
 &=& -I_{z,m}|_{\mbox{\tiny R}}
\label{eq9}
\end{eqnarray} 
From the above mathematical argument the sign reversal of $I_{z,m}$ under
interchanging the roles played by $\alpha$ and $\beta$ can be easily 
understood. Certainly this vanishing behavior can be exploited to determine
any one among these two SO fields if the other is given. In particular the
determination of Dresselhaus strength will be much easier as for a specific
material it is constant, while Rashba strength can be tuned with the help
of external gate potential. Here, it is worthy to note that the interplay 
of Rashba and
Dresselhaus SO couplings has also been reported in several other contexts,
and particularly when these two strengths are equal, persistent spin 
helix~\cite{sh1,sh2,sh3,sh4} has observed which is of course one of the 
most important and attractive areas of spintronics.

\begin{figure}[ht]
{\centering \resizebox*{8cm}{8cm}{\includegraphics{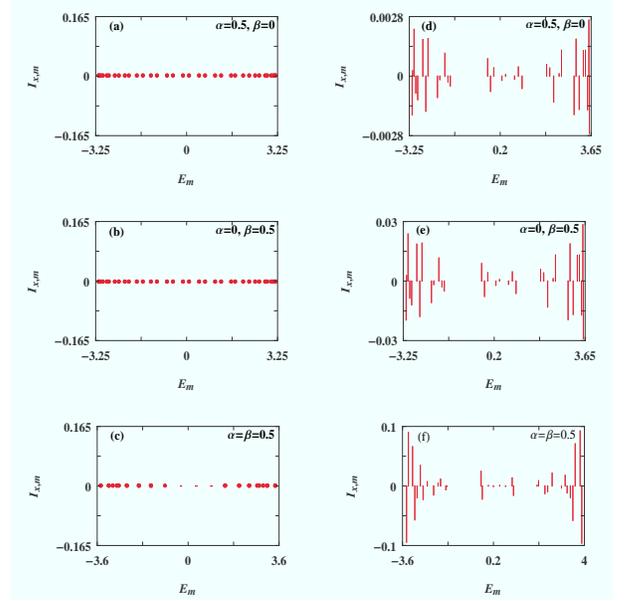}}\par}
\caption{(Color online). $X$-component of spin current ($I_{x,m}$) carried
by distinct energy levels of a mesoscopic ring where different spectra 
correspond to the identical meanings as described in Fig.~\ref{fig2a}. The 
ring size and other physical parameters are also same as taken in 
Fig.~\ref{fig2a}.}
\label{fig2b}
\end{figure}

\begin{figure}[ht]
{\centering \resizebox*{8cm}{8cm}{\includegraphics{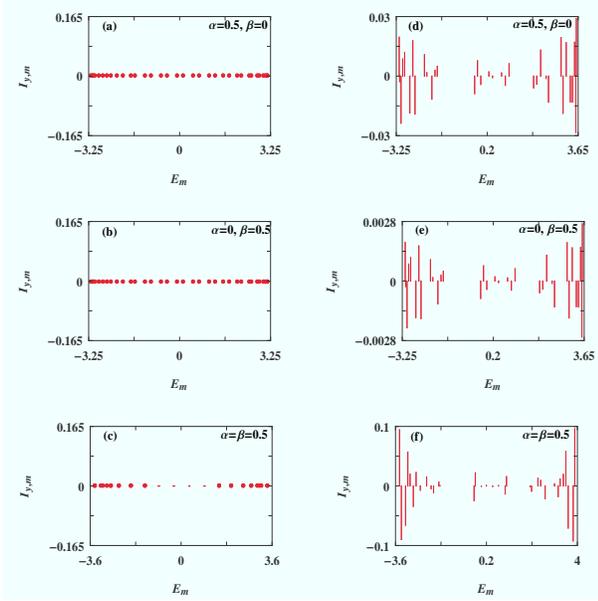}}\par}
\caption{(Color online). $Y$-component of persistent spin current ($I_{y,m}$)
for separate energy levels of a conducting ring in presence of $\alpha$ and
$\beta$, where the different spectra represent the similar meanings as 
described in Fig.~\ref{fig2a}. The physical parameters remain unchanged as
taken in Fig.~\ref{fig2a}.}
\label{fig2c}
\end{figure}

\begin{figure}[ht]
{\centering \resizebox*{6cm}{5cm}{\includegraphics{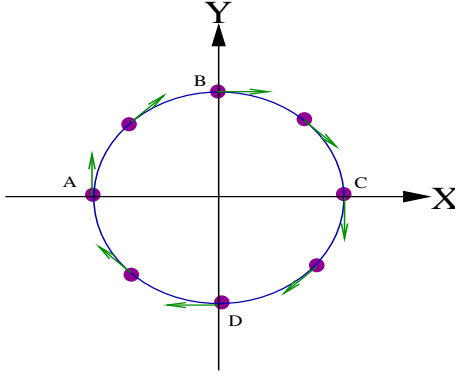}}\par}
\caption{(Color online). Velocity direction (green arrow) of a moving 
electron at some typical points (filled colored circles) of a ring placed 
in the $X$-$Y$ plane.}
\label{fig2aa}
\end{figure}

In Figs.~\ref{fig2b} and \ref{fig2c} we present the characteristics of $X$
and $Y$ components of spin current, respectively, carried by individual 
energy levels for the identical ring size and parameter values as taken in
Fig.~\ref{fig2a} for finer comparison of all three components. The 
observations are noteworthy. (i) For the perfect ring (viz, $\epsilon_A=
\epsilon_B=0$), both $X$ and $Y$ components of current are zero for each 
energy eigenstates, while non-zero contribution comes from the 
Fibonacci ring. In order to explain this behavior let us focus on 
Fig.~\ref{fig2aa}, where the velocity direction of a moving electron is 
schematically shown at different lattice points of a ring placed in the 
$X$-$Y$ plane. Now, consider the Rashba and Dresselhaus Hamiltonians in 
a continuum representation where they get the forms: 
$\mbox{\boldmath $H$}_{\mbox{\tiny R}} = \alpha 
\left(\mbox{\boldmath $\sigma$}_y \mbox{\boldmath $p$}_x -
\mbox{\boldmath $\sigma$}_x \mbox{\boldmath $p$}_y\right)$
and
$\mbox{\boldmath $H$}_{\mbox{\tiny D}} = \beta 
\left(\mbox{\boldmath $\sigma$}_y \mbox{\boldmath $p$}_y -
\mbox{\boldmath $\sigma$}_x \mbox{\boldmath $p$}_x\right)$
where $\mbox{\boldmath $p$}_x$ and $\mbox{\boldmath $p$}_y$ are the
components of $\mbox{\boldmath $p$}$ along $X$ and $Y$ directions,
respectively. Thus, at the point A of a pure Rashba ring 
only $\mbox{\boldmath $p$}_y$ will contribute (since here 
$\mbox{\boldmath $p$}_x=0$) to $\mbox{\boldmath $H$}_{\mbox{\tiny R}}$, 
while it is 
$-\mbox{\boldmath $p$}_y$ at the point C (Fig.~\ref{fig2aa}). Similarly, 
for the points B and D the contributing terms are $\mbox{\boldmath $p$}_x$ 
and $-\mbox{\boldmath $p$}_x$, respectively. As a result of this the net 
contribution to $\mbox{\boldmath $H$}_{\mbox{\tiny R}}$ becomes zero, and
this is equally true for any other diagonally opposite points (though in
these cases both $\mbox{\boldmath $p$}_x$ and $\mbox{\boldmath $p$}_y$
contribute) which leads to a
vanishing spin current along $X$ and $Y$ directions for a perfect Rashba
ring. The same argument is also valid in the case of a pure Dresselhaus 
ring. But, as long as the symmetry between the diagonally opposite points 
is broken the mutual cancellation does not take place which results a finite
non-zero spin current along these two directions. This is exactly what we 
get in a Fibonacci ring, and in the same footing, we can expect non-zero
spin current for any other disordered rings.
(ii) In the Fibonacci ring the $X$-component ($Y$-component) of spin current 
in presence of $\alpha$ maps exactly in the opposite sense (i.e., identical
magnitude but opposite in direction) to the $Y$-component ($X$-component)
of current under swapping the roles played by $\alpha$ and $\beta$ (right
columns of Figs.~\ref{fig2b} and \ref{fig2c}). This phenomenon can be 
explained from the following mathematical analysis.
\begin{eqnarray}
I_{y,m}|_{\mbox{\tiny D}} & = &
\langle\psi^{\prime}_m|\mbox{\boldmath $I$}_{y,m}|\psi^{\prime}_m\rangle
\nonumber \\
 &=& \langle\psi_m|\mbox{\boldmath $U$}^\dagger\mbox{\boldmath
$I$}_{y,m}\mbox{\boldmath $U$}|\psi_m\rangle\nonumber \\
 &=& \langle\psi_m|\mbox{\boldmath $U$}^\dagger\frac{1}{2N}
\left(\mbox{\boldmath$\sigma$}_y\dot{\mbox{\boldmath $x$}} + 
\dot{\mbox{\boldmath $x$}} \mbox{\boldmath $\sigma$}_y\right)
\mbox{\boldmath $U$}|\psi_m\rangle\nonumber \\
 &=& \langle\psi_m|\frac{1}{2N}
\left(-\mbox{\boldmath$\sigma$}_x\dot{\mbox{\boldmath $x$}} - 
\dot{\mbox{\boldmath $x$}}\mbox{\boldmath $\sigma$}_x\right)|\psi_m\rangle
\nonumber \\
 &=& -I_{x,m}|_{\mbox{\tiny R}}
\label{eq10}
\end{eqnarray}
and
\begin{eqnarray}
I_{x,m}|_{\mbox{\tiny D}} & = &
\langle\psi^{\prime}_m|\mbox{\boldmath $I$}_{x,m}|\psi^{\prime}_m\rangle
\nonumber \\
 &=& \langle\psi_m|\mbox{\boldmath $U$}^\dagger\mbox{\boldmath$I$}_{x,m}
\mbox{\boldmath $U$}|\psi_m\rangle\nonumber \\
 &=& \langle\psi_p|\mbox{\boldmath $U$}^\dagger\frac{1}{2N}
\left(\mbox{\boldmath$\sigma$}_x\dot{\mbox{\boldmath $x$}} + 
\dot{\mbox{\boldmath $x$}}\mbox{\boldmath $\sigma$}_x\right)
\mbox{\boldmath $U$}|\psi_m\rangle\nonumber \\
 &=& \langle\psi_m|\frac{1}{2N}
\left(-\mbox{\boldmath$\sigma$}_y\dot{\mbox{\boldmath $x$}} - 
\dot{\mbox{\boldmath $x$}}\mbox{\boldmath $\sigma$}_y\right)|\psi_m\rangle
\nonumber \\
 &=& -I_{y,m}|_{\mbox{\tiny R}}
\label{eq11}
\end{eqnarray}
Equations~\ref{eq10} and \ref{eq11} clearly describe the interchange of $X$ 
and $Y$ components of current under the reciprocation of $\alpha$ and $\beta$.
(iii) Quite interestingly we see that both for these two components 
($X$ and $Y$) the states lying towards the energy band edge carry higher 
current compared to the inner states, unlike the $Z$-component of current
where opposite signature is noticed. This feature is also observed in other 
quasi-periodic rings as well as in a fully random one. In addition, a 
significant change in current amplitude takes place between the two 
current components when the ring is subjected to either $\alpha$ or 
$\beta$, even if these strengths are identical (Figs.~\ref{fig2b}(d) and 
(e); (Figs.~\ref{fig2c}(d) and (e)), though its proper physical explanation
is not clear to us.
(iv) Finally, it is important to note that at $\alpha=\beta$ none of these
$X$ and $Y$ components of current vanishes (see Figs.~\ref{fig2b}(f) and
\ref{fig2c}(f)) since 
$I_{y,m}|_{\mbox{\tiny D}} \ne -I_{y,m}|_{\mbox{\tiny R}}$ and also
$I_{x,m}|_{\mbox{\tiny D}} \ne -I_{x,m}|_{\mbox{\tiny R}}$.

\subsection{Components of net current for a particular electron filling}

Now we focus on the behavior of all three components of spin current for 
a particular electron filling and the total spin current taking the 
contributions from these individual components.

\begin{figure}[ht]
{\centering \resizebox*{7cm}{7cm}{\includegraphics{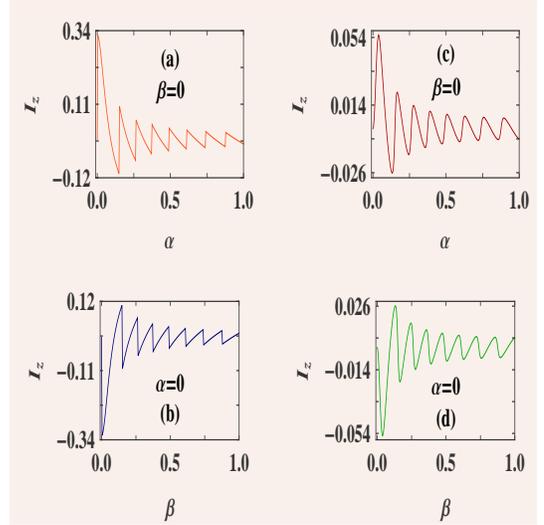}}\par}
\caption{(Color online). $Z$-component of spin current ($I_z$) as a function
of any one of two SO fields (keeping the other at zero) for a $89$-site ring
($10$th generation) considering $N_e=50$, where the left and right columns 
correspond to $\epsilon_A=\epsilon_B=0$ and $\epsilon_A=-\epsilon_B=1$,
respectively.}
\label{fig3a}
\end{figure}

In Fig.~\ref{fig3a} we present the variation of $Z$-component of spin 
current as a function of SO coupling for a $89$-site ring considering
$N_e=50$. In the first column the results are shown for a perfect ring,
while for the Fibonacci ring they are presented in the other column.
The current exhibits an anomalous oscillation with SO coupling and its 
amplitude gradually decreases with increasing the coupling strength.
This oscillation is characterized by the crossing of different distinct
energy levels (viz, degeneracy) of the system, and also observed in other 
context~\cite{san1,site3}.  
The other feature i.e., the phase reversal of $I_z$ by interchanging 
the parameters $\alpha$ and $\beta$ can be well understood from our
\begin{figure}[ht]
{\centering \resizebox*{7cm}{7cm}{\includegraphics{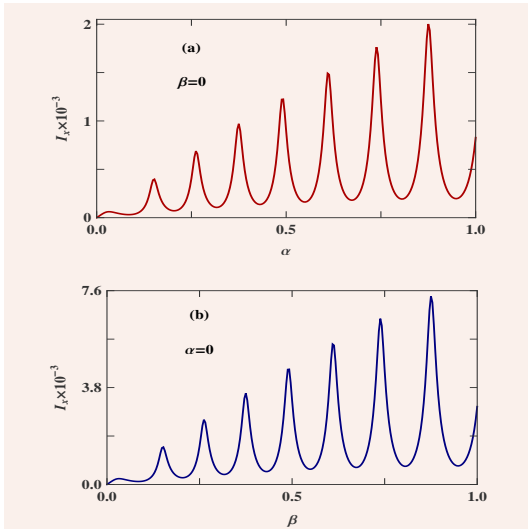}}\par}
\caption{(Color online). $X$-component of spin current ($I_x$) as a function
of any one of the two SO interactions for a $89$-site ($10$th generation) 
Fibonacci ring ($\epsilon_A=-\epsilon_B=1$) considering $N_e=50$.}
\label{fig3b}
\end{figure}
\begin{figure}[ht]
{\centering \resizebox*{7cm}{7cm}{\includegraphics{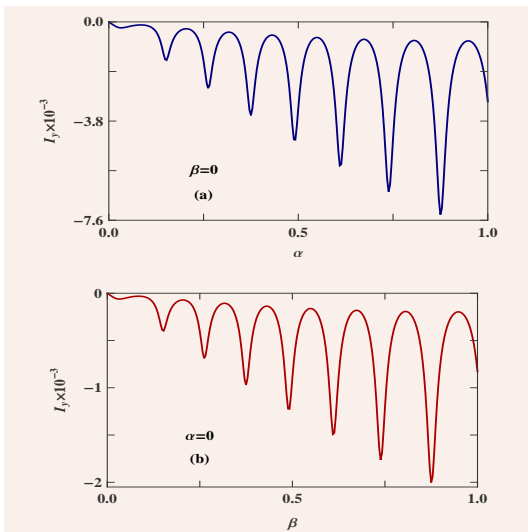}}\par}
\caption{(Color online). $Y$-component of spin current ($I_y$) as a function
of any one of the two SO interactions for a $89$-site ($10$th generation) 
Fibonacci ring ($\epsilon_A=-\epsilon_B=1$) considering $N_e=50$.}
\label{fig3c}
\end{figure}
previous analysis, and thus, the net $Z$-component of spin current 
should vanish under the situation $\alpha=\beta$, which is not shown
here to save space. In addition, it is also observed that the net
current amplitude in the Fibonacci ring (right column of Fig.~\ref{fig3a}) 
for any $\alpha$ or $\beta$ is much smaller than the perfect one (left
column of Fig.~\ref{fig3a}), and it is solely associated with
the conducting nature of different energy levels those are contributing
to the current. The nature of current carrying states can be clearly noticed
from the spectra given in Fig.~\ref{fig2a}, where the currents carried by
distinct energy levels of the Fibonacci ring are much smaller compared to
the perfect one.

The behaviors of other two current components (viz, $X$ and $Y$) are shown
in Figs.~\ref{fig3b} and \ref{fig3c}. Since both these two components are
zero for the perfect ring, here we present the results only for a Fibonacci
ring considering the identical parameter values and electron filling as 
taken in Fig.~\ref{fig3a}. The current exhibits an oscillation, and 
unlike $Z$-component, the oscillating peak increases with increasing 
SO interaction. 

\begin{figure}[ht]
{\centering \resizebox*{8cm}{6.5cm}{\includegraphics{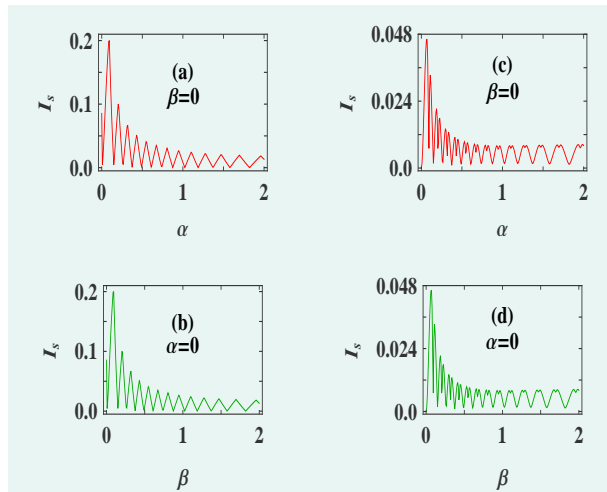}}\par}
\caption{(Color online). Net spin current $I_s$ (considering the 
contributions from all three components) as a function of any one of two
SO interactions (setting the other at zero) for a $89$-site ($10$th 
generation) ring with $50$ electrons, where the first and second columns 
correspond to $\epsilon_A=\epsilon_B=0$ and $\epsilon_A=-\epsilon_B=1$, 
respectively.}
\label{itotal}
\end{figure}

Finally, focus on the spectra shown in Fig.~\ref{itotal} where we present
the variation of net spin current $I_s$ taking the individual contributions
from three different components. It is defined as 
$I_s=\sqrt{I_x^2+I_y^2+I_z^2}$. Both the perfect and Fibonacci rings
are taken into account those are placed in the first and second columns
of Fig.~\ref{itotal}, respectively. In both these two cases we find 
oscillating behavior of current following the current components as discussed
in Figs.~\ref{fig3a}-\ref{fig3c}. For the ordered ring since the contribution
comes only from $I_z$, the oscillation of $I_s$ gradually dies out with SO
coupling. While for the Fibonacci ring as $I_x$ and $I_y$ along with $I_z$
contribute to $I_s$, a finite but small oscillation still persists even for
higher values of SO coupling. The another feature obtained from the spectra
i.e., lesser $I_s$ in Fibonacci ring compared to the perfect one for any
non-zero SO coupling is quite obvious.  

\subsection{Prediction of on-site energy}

In this sub-section we discuss the possibilities of estimating any one of
the two on-site potentials ($\epsilon_A$ and $\epsilon_B$) of a Fibonacci 
ring if we known the other one. 

This can be done quite easily by analyzing the behavior of current amplitude
of individual components as a function of either $\epsilon_A$ or $\epsilon_B$,
setting the other constant, keeping in mind that a distinct feature may
appear when these two site energies become identical since the current 
components are significantly influenced by the disorderness.

\begin{figure}[ht]
{\centering \resizebox*{7cm}{4cm}{\includegraphics{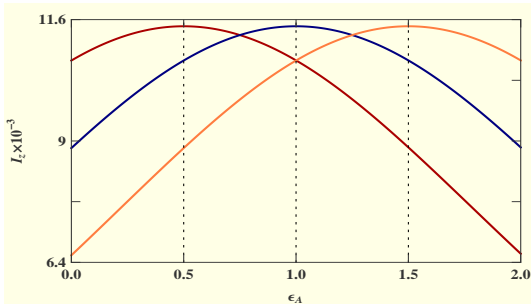}}\par}
\caption{(Color online). $I_z$ vs. $\epsilon_A$ for a $55$-site ($9$th 
generation) Fibonacci ring for three distinct
values of $\epsilon_B$, where the red, navy and orange curves correspond to
$\epsilon_B=0.5$, $1$ and $1.5$, respectively. The other physical parameters
are: $N_e=55$, $\alpha=1$ and $\beta=0$.}
\label{fig4a}
\end{figure}

In Fig.~\ref{fig4a} we show the variation of $I_z$ as a function of 
\begin{figure}[ht]
{\centering \resizebox*{7cm}{4cm}{\includegraphics{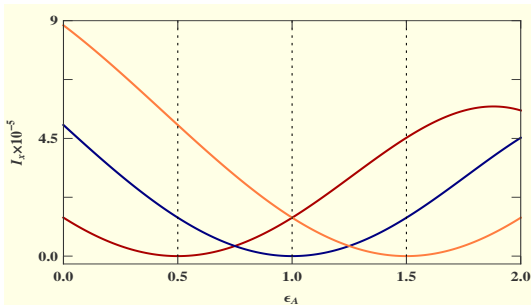}}\par}
\caption{(Color online). Same as Fig.~\ref{fig4a} where the variation of
$I_x$ with $\epsilon_A$ is shown.}
\label{fig4b}
\end{figure}
$\epsilon_A$ for a $55$-site Fibonacci ring with $55$ electrons considering 
three distinct values of $\epsilon_B$ those are represented by three 
different colored curves. Here we take $\alpha=1$ and fix $\beta$ to zero.
Interestingly we see that $I_z$ reaches to the maximum (shown by the dotted
line) when the two site energies are equal. While, the current amplitude 
gets reduced with increasing the deviation of site energies i.e., 
$|\epsilon_A-\epsilon_B|$. This is solely associated with localizing behavior
of electronic waves and directly linked with previous analysis. Thus, for a
particular material composed of two different lattices one can determine the 
site energy of any one by varying the other and observing the maximum of 
$I_z$. It takes place only when $\epsilon_A=\epsilon_B$.

In the same footing, we can also find a definite condition from the other 
two current components through which site energy is predicted. The results
\begin{figure}[ht]
{\centering \resizebox*{7cm}{4cm}{\includegraphics{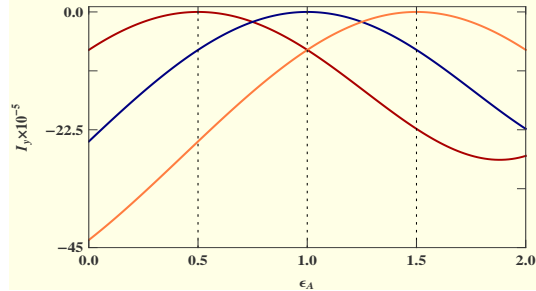}}\par}
\caption{(Color online). Same as Fig.~\ref{fig4a} where the variation of
$I_y$ with $\epsilon_A$ is shown.}
\label{fig4c}
\end{figure}
are presented in Figs.~\ref{fig4b} and \ref{fig4c} for the identical ring
and parameter values as taken in Fig.~\ref{fig4a}. From these spectra it
is clearly seen that when $\epsilon_A$ becomes equal to $\epsilon_B$, both
$X$ and $Y$ components of spin current drop exactly zero, and this vanishing 
behavior leads to the possibility of determining site energy.

Before we end this sub-section it is important to note that from the 
practical point of view one may think how site energies of a large number
of $A$-type or $B$-type sites can be tuned for large sized ring. This is 
of course a difficult task. But, through the present analysis we intend 
to establish that if we take a ring geometry with few foreign atoms, then 
this prescription will be useful since tuning the site energies of only 
these few atoms by means of external gate potential site energies of 
parent atoms can be estimated.  

\section{Summary and conclusions}

In summary, we have made a comprehensive analysis of all three components
of persistent spin current in a quasi-periodic Fibonacci ring with Rashba
and Dresselhaus SO interactions. Within a tight-binding framework we compute
all the current components based on second-quantized approach. Several 
distinct features have been observed those can be utilized to determine
any one of the two SO fields as well as the site energies when the other 
is known.

The results studied in this work are for a generic model, not related to
any specific material, and thus, can be extended to any such correlated
as well as uncorrelated lattice models and can provide some basic inputs
towards spin dependent transport phenomena.

\section{Acknowledgment}

MP is thankful to University Grants Commission (UGC), India for
research fellowship.

\end{document}